\documentclass[12pt,a4paper]{article}%
\pdfoutput=1
\usepackage{amsmath,amssymb,amsfonts,wasysym}
\usepackage[pdftex]{graphicx}
\usepackage{color}
\usepackage{float}
\usepackage[bf,footnotesize]{caption2}
\usepackage[]{hyperref}%
\numberwithin{equation}{section}
\setcaptionmargin{1cm}

\def\({\left(}
\def\){\right)}

\def\be{\begin{equation}}
\def\ee{\end{equation}}

\begin{document}
\begin{flushright}
IFUP-TH/2010-18
\end{flushright}
\vskip 1.0cm
\begin{flushleft}
\small Based on a talk given at the conference\\
IFAE2010 - Incontri di Fisica delle Alte Energie\\
``Sapienza'' University, Rome, 7-9 April 2010
\end{flushleft}
\vskip 1.0cm
\begin{center}
{\Large \bf Associated scalar-vector production at the LHC within an effective Lagrangian approach} \vskip 1.0cm
{\large Riccardo Torre}\\[0.3cm]
{\it Universit\`a di Pisa and INFN Sezione di Pisa, Dipartimento di Fisica, Largo Fibonacci 3, I-56127 Pisa, Italy}\\[5mm]
\end{center}
\vskip 1.0cm
\begin{abstract} 
We consider the case in which a strong dynamics is responsible for Electro-Weak Symmetry Breaking (EWSB) and both a scalar $h$ and a  vector $V$, respectively a singlet and a triplet under a custodial $SU\left(2\right)$, are relevant and have a mass below the cut-off $\Lambda\approx 4\pi v$. In this framework we study the total cross sections for the associated $Vh$ production at the LHC at $14$ TeV as functions of two independent free parameters.
\end{abstract}


\section{Introduction}
The LHC will be able to explore the scale of Electro-Weak Symmetry Breaking (EWSB), i.e. the Fermi scale $v=246$ GeV hopefully shedding light on the mechanism that generates it. 
If a strong dynamics is responsible for the EWSB, new degrees of freedom should become relevant at the Fermi scale in order to take under control the asymptotic behavior of the amplitudes for the longitudinal gauge boson scattering.
In the framework of a strongly interacting dynamics for EWSB, we are interested in the case in which a scalar $h$ and a vector $V$, respectively a singlet and a triplet under a custodial $SU\(2\)$ with a mass below the cut-off $\Lambda \approx 4\pi v$, share the task of unitarizing the $W_{L}W_{L}\to W_{L}W_{L}$ scattering. In particular we want to study the phenomenology of the $Vh$ associated production at the LHC since it could be the main signature of the spectrum that we are considering.

\section{One vector and one scalar below the cut-off}\label{sec2}
We would like to construct a model-independent Lagrangian to describe the new degrees of freedom $h$ and $V$ without making any hypotheses on the origin of the light scalar: it could be a Strongly Interacting Light Higgs (SILH) boson in the sense of \cite{Contino:2010} or a more complicated object arising from an unknown strong dynamics. Nevertheless, in order to construct a phenomenologically relevant Lagrangian we have to make some assumptions that can be stated as follows:

\begin{enumerate}
\item Before weak gauging, the Lagrangian responsible for EWSB has a $SU\(2\)_{L}\times SU\(2\)^{N}\times SU\(2\)_{R}$ global symmetry, with $SU\(2\)^{N}$ gauged, spontaneously broken to the diagonal $SU\(2\)_{d}$ by a generic non-linear sigma model;
\item Only one vector triplet $V_{\mu}^{a}$ of the $SU\(2\)^{N}$ gauge group has a mass below the cut-off $\Lambda\approx 3$ TeV, while all the other heavy vectors can be integrated out. Furthermore the new vector triplet $V_{\mu}^{a}$ couples to fermions only through the mixing with the weak gauge bosons of $SU\(2\)_{L}\times U\(1\)_{Y}$ ($Y=T_{3R}+1/2\(B-L\)$);
\item The spectrum also contains a light scalar singlet of $SU\(2\)_{d}$ with a relatively low mass $m_{h}\leq v$. 
\end{enumerate}

With these assumptions we are able to construct an effective Lagrangian (see \cite{Carcamo:2010} and \cite{Barbieri:2010}) with a cut-off $\Lambda\approx 3$ TeV to study the associated $Vh$ production at the LHC. The general Lagrangian depends in principle on five couplings and on the scalar and vector masses, $m_{h}$ and $M_{V}$ respectively. The five couplings can be related each other by requiring unitarity of the two body $W_{L}W_{L}$ scattering amplitudes. It is simple to show that requiring a constant asymptotic behavior at least for the elastic channel ($W_{L}W_{L}\to W_{L}W_{L}$) the parameter space can be reduced to only two independent parameters for fixed values of $m_{h}$ and $M_{V}$ \cite{Carcamo:2010}. We can choose as the independent parameters sets either the pairs $\(G_{V},d\)$ or $\(a,d\)$ where $G_{V}$, $a$ and $d$ are the couplings of the vector $V$ and the scalar $h$ to the weak Goldstone bosons and the coupling of $hVV$ respectively. 

\section{Associated $Vh$ production at the LHC}
\begin{figure}[!htb]
\caption{Total cross sections for the DY associated $Vh$ productions as functions of the heavy vector mass at the LHC for $\sqrt{s}=14$ TeV, $m_{h}=180$ GeV, $d=1$ and for different values of $G_{V}$.}
\centering
\vspace{5mm}\includegraphics[width=8.5cm]{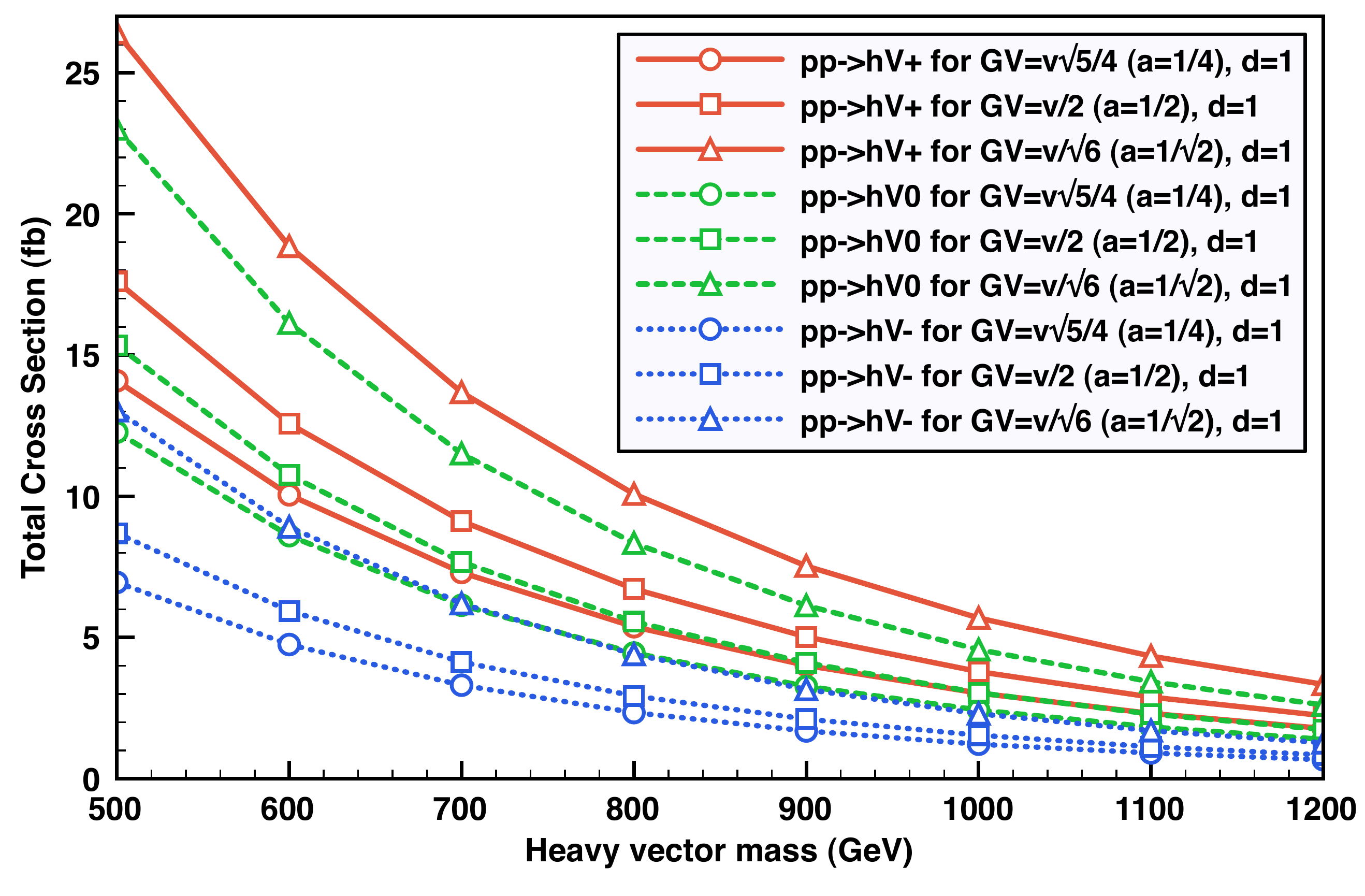}
\vspace{-4mm}
\label{fig1}
\end{figure}

The most relevant channel for the associated $Vh$ production at the LHC is the Drell-Yan (DY) annihilation. There are three different charge configuration for the $Vh$ system: $hV^{-}$, $hV^{0}$ and $hV^{+}$. The DY total cross sections at the LHC at $14$ TeV as functions of the heavy vector mass $M_{V}$ for $m_{h}=180$ GeV and for different values of the relevant couplings are depicted in Figure \ref{fig1}. From this Figure we see that the DY total cross sections are of order of $10$ fb for a reference value $d=1$. 
The order of magnitude of the total cross sections for the DY associated production is comparable 
with the corresponding values of the total cross sections for the $VV$ and $hh$ productions\footnote{For the $hh$ and $VV$ pair productions at the LHC see respectively \cite{Contino:2010} and \cite{Barbieri:2010}.}. 
Moreover, since the DY associated production is generated only through the $hVV$ interaction, the total cross section is proportional to $d^{2}$ and therefore the total cross sections of Figure \ref{fig1} are enhanced by a factor $d^{2}$ for $d>1$. 

\section{Multi-lepton events}
Assuming $BR\(h\to W^{+}W^{-}\)\approx 1$, with the values of the total cross sections shown in Figure \ref{fig1}  we can compute the expected number of same-sign di-lepton and tri-lepton events for a reference integrated luminosity $L=100$ fb$^{-1}$. For $M_{V}=700$ GeV we find the number of multi-lepton events shown in Table \ref{table1}.

\begin{table}[htb!]
\caption{Total number of same sign di-lepton and tri-lepton events ($e$ or $\mu$ from $W$ decays) for the DY associated $Vh$ production at the LHC for $\sqrt{s}=14$ TeV, $L=100$ fb$^{-1}$, $M_{V}=700$ GeV and $m_{h}=180$ GeV, $d=1$ and for different values of the parameter $G_{V}$.}
\centering
\begin{tabular}[c]{|c|c|c|c|}
		\hline
		$G_{V}$ & $a$ & di-leptons & tri-leptons \\
		\hline
		$\sqrt{5}v/4$ & $1/4$ & $102.4$ & $30.3$ \\
		\hline
		$v/2$  & $1/2$ & $128.0$ & $37.8$ \\
		\hline
		$v/\sqrt{6}$  & $1/\sqrt{2}$ & $192.0$  & $56.7$ \\
		\hline
\end{tabular}\label{table1}
\end{table}

\section{Summary and conclusions}
We have considered the case in which a scalar-vector system is relevant in a strongly interacting framework for EWSB. We have used a model-independent approach to construct an effective Lagrangian to study the associated $Vh$ production at the LHC. The number of multi-lepton events in the high luminosity phase of the LHC is of order of $100$. A careful study of the background, that is beyond our scope, should be done to see if these events could emerge from it.

\section{Acknowledgments}
This work was done in collaboration with Antonio E. Carcamo Hernandez. \\
I would like to thank Riccardo Barbieri for many useful suggestions and Enrico Trincherini for suggesting us the main idea of studying the associated production of a light scalar and a heavy vector. I also thank Gennaro Corcella, Slava Rychkov and Riccardo Rattazzi for useful discussions.

\end{document}